\documentclass[conference]{IEEEtran}
\IEEEoverridecommandlockouts
\usepackage{cite}
\usepackage{amsmath,amssymb,amsfonts}
\usepackage{algorithmic}
\usepackage{graphicx}
\usepackage{subfigure}
\usepackage{soul}
\usepackage{textcomp}
\usepackage{xcolor}
\usepackage{authblk}
\usepackage{amsthm}

\theoremstyle{plain}

\begin{document}

\title{Jamming-Aware Control Plane in Elastic Optical Networks}
\author[1]{\'Italo Brasileiro} 
\author[2]{Mounir Bensalem} 
\author[1]{Andr\'e Drummond} 
\author[2]{Admela~Jukan} 
\affil[1]{University of Brazilia, Brazil}
\affil[2]{Technische Universit\"at Braunschweig, Germany}


\maketitle
\begin{abstract}
Physical layer security is essential in optical networks. In this paper, we study a jamming-aware control plane, in which a high power jamming attack exists in the network. The studied control plane considers that the jammed connections  can be detected and avoided. We used a physical layer model, in which we embedded the additional jamming power, to evaluate different security in scenarios,  such as a jamming-free scenario, jamming with an unaware controller, and jamming with an aware controller. The performance is analyzed in terms of the blocking rate and slots utilization. We analyze the impact of jamming attacks in the least used link and in the most used link on the network. The results demonstrates that the jamming avoidance by the control plane can reach performance near the not jammed scenario.
\end{abstract}
%
\section{Introduction} 
The development of new applications with high bandwidth demand and new access technologies pushes the telecommunications industry to seek new solutions to attend the growing need for bandwidth \cite{lopez:2018}. Elastic optical networks (EON) appear in this context as an infrastructure candidate to accomplish the requirement for higher transmission rates \cite{chatterjee:2015}. The EON links present channels called slots,  with the spectral granularity of 12.5 GHz. The spectral slots can be grouped and create larger channels to provide higher transmission rates to greater bandwidth. The data transmission in the optical environment, the establishment of circuits in parallel on the same link and the flexibility to create channels that can settle higher bandwidth circuits are important characteristics of EONs that increase the transmission capacity \cite{Chen:15}.

To enable the use of EON in the real world, it is convenient to ensure a high level of security on the network. In general, EON can be the target of several types of attacks \cite{patel:2012}. One particular type of attack is performed by inserting a high power signal in a spectral frequency range within the optical fiber. This power addition is called jamming attack and leads to the reduction of the signal-to-noise ratio (SNR) for the active circuits. As a consequence of the SNR reduction, the circuits may reach SNR rates below a defined SNR threshold ($SNR_{TH}$), which impairs the quality of transmission (QoT) of the network and can cause financial loss for network operators, by the breach of guarantees defined in the Service Level Agreement (SLA).

\begin{figure}[!h]
    \begin{center}
        \includegraphics[width=0.45\textwidth]{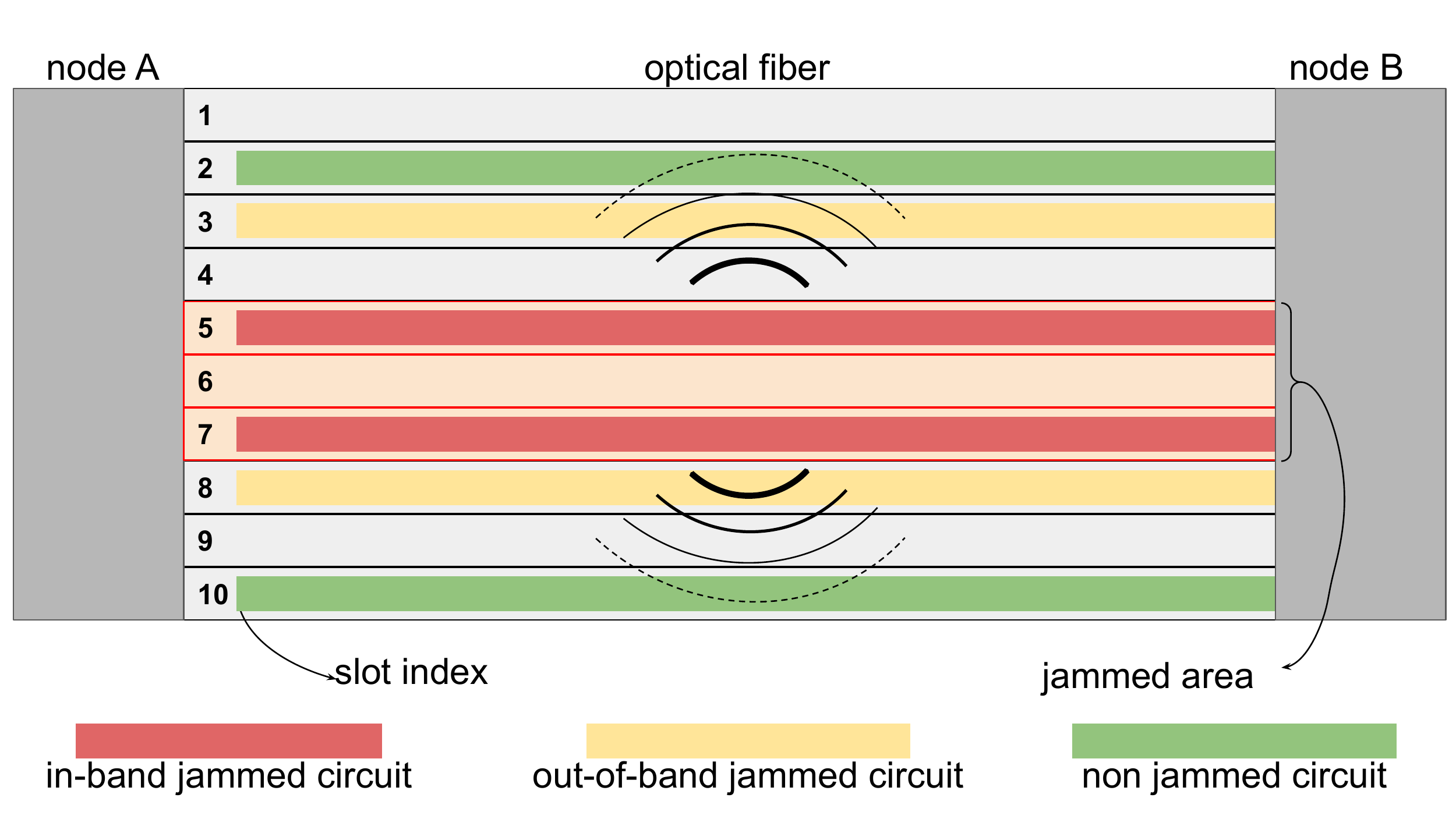}
        \caption{\textit{In-band} and \textit{out-of-band} jamming demonstration.}
        \label{fig:jammed}
    \end{center}
\end{figure}

Jamming attacks can be classified into two types \cite{medard1998attack}. Figure \ref{fig:jammed} shows an optical link between two nodes, under a jamming attack. The first type is \textit{in-band jamming}, and represents the insertion of high power within a spectral window (slots 5, 6 and 7 in Fig. \ref{fig:jammed}). The slots inside the affected range undergo an increase in power, which directly impacts their SNR level (circuits in slots 5 and 7 in Fig. \ref{fig:jammed}). The second type of attack, out-of-band jamming, is a side-effect of the interference provided by the power of jammed circuits in the other circuits of the network (in slots 3 and 8). The out-of-band intensity is stronger in the slots near the spectral range under attack. Thus, the jamming affects circuits in the attacked slot range and also circuits outside the attack area.

The study of physical interference requires a model to measure the impact of the physical transmission environment on the quality of network circuits. Most papers focuses on the proposition of solutions and models for detecting jamming attacks \cite{bensalem2019detecting, natalino2018field, li2018light} or protecting the network against its occurrence \cite{prucnal2009physical, hu2015chaos,li2016fast,singh2017combined}. However, it is important to analyze the non-linearity effects added by the insertion of a high power signal. The authors in \cite{johannisson2014modeling} propose a model for calculating the quality of the circuits, using their respective SNR. The power of jamming attack can be added in the SNR model, and the proposal in \cite{johannisson2014modeling} can be modified to consider the impact caused by the jamming, both in the circuits that are under attack and in the circuits allocated outside the jammed area.

This paper presents a performance evaluation for a control plane aware of jamming attacks. Through signal processing at the circuit destination node, the control panel is informed about the presence of circuits with power affected by the jamming attack. With this information, the control plane prevents the occurrence of circuits that are targets of jamming, preventing the propagation of the effect in other circuits of the network. The evaluation of the control plane with jamming avoidance is made in comparison with a control plane unaware of jamming attacks, under different scenarios of jamming attacks. The results show that avoiding the establishment of attacked circuits in some cases can generate performance similar to scenarios without attack.

The rest of this paper is organized as follows. Section \ref{sec:jamming} shows the proposed jamming model for EON. Section \ref{sec:detection} shows the proposed jamming avoidance model scheme. Section \ref{sec:evaluation} presents the performance evaluation for the proposed model. Finally, we conclude the paper in Section \ref{sec:conclusion}.

\section{Jamming-Aware Physical Layer Model in EON}\label{sec:jamming}
We adopt the  physical layer model proposed in \cite{bensalem2020embedding}, where jamming attacks are embedded in the SNR equation \cite{johannisson2014modeling}.
We model the network infrastructure as a graph $G=(V,E)$.  The set of nodes is represented by $V$, and  consists of routing devices, and transceivers., and the set of optical links by $E$.
 An optical link contains is a set of connected fiber spans, such that each span is an optical fiber followed by an erbium-doped fiber amplifier (EDFA). A link is able to transmit M slots. We define the center frequency $f_m$, the bandwidth $\Delta f_m$, and the power $P_m$, for each channel $m=1,...,M$.  We consider a network that can support an large number of connections established between different nodes and following a predefined route. A connection $i$ can be associated to a route $r_i$, where $r_i$ is a set of links starting from the source node and ending by the destination node of connection $i$.
%

%
%
A jammer $J$ can insert a high power signal in one or more channels in order to disrupt the service. We denote by $M_J \subset M$ the set of jammed channels. We assume that the power of $J$ is $P_J= P + \epsilon$, where $P$ is the power of the rest of channels, and $\epsilon$ is the additional power. \\

The jamming attack aware SNR equation associated to a connection $i$ using a route $r_i$ is expressed as follows:

\begin{equation}\label{eq:snr}
    SNR_m = \frac{G_m}{G^{ASE} + G^{NLI,s} + G^{J}}
\end{equation}
Where $G_m$ is the signal power spectral density (PSD) of connection $i$ using a channel $m$, $G^{ASE}$ is the PSD of the amplified spontaneous emission (ASE) noise, $G^{NLI,s}$ is the PSD of the noise from nonlinear impairments (NLI) in a secure network, and $G^J$ is the PSD of the noise from NLI caused by jamming.\\

The PSD of connection $i$ using a channel $m$ is written as:
\begin{equation}\label{PSD}
    G_m = \frac{P_m}{\Delta_m}
\end{equation}
Where $P_m$ is the power corresponding to $G_m$, and $\Delta_m$ denotes the bandwidth and considered constant for all links in the connection.\\
The PSD of ASE can be expressed as follows:
\begin{equation}
    G^{ASE}=\sum_{l\in r_i}N_l G_{0}^{ASE},\;\;\;\; G_{0}^{ASE}=(e^{\alpha L} - 1)Fh\nu
\end{equation}
Where $N_l$ is the number of spans on link $l$, $L$ is the length of each span, $\alpha$ is the power attenuation, $F$ is the spontaneous emission factor, $h$ is Planck’s constant, and $\nu$ is the light frequency.\\

The PSD of the noise from NLI in a secure network:
\begin{equation}
    G^{NLI, s}=\sum_{l\in r_i}N_l G_{l}^{NLI, s}
\end{equation}
with,
\begin{multline}
    G_{l}^{NLI, s}(f_m) = \phi G_m [G_m^2 \text{arcsinh}(\rho (\Delta f_m)^2) \\ + \sum_{\substack{m' = 1 \\ m' \neq m}}^{M}G_{m'}^2 \text{ln}(\frac{f_{m,m'} + \Delta f_{m'}/2}{f_{m,m'} - \Delta f_{m'}/2}) ]
\end{multline}
and,
\begin{equation}
    \phi = \frac{3\gamma^2}{2 \pi \alpha |\beta_2|}, \;\;\;\;\; \rho = \frac{\pi^2 |\beta_2|}{2\alpha}
\end{equation}
Where $f_m$ is the center frequency of channel $m$.  $m'$ is another connection using link $l$, $\Delta f_{m'}$ is the bandwidths for connection $m'$, $f_{m,m'}=|f_m - f_{m'}|$ is the center frequency spacing between connections $m$ and $m'$, $\gamma$ is the fiber nonlinearity coefficient, and $\beta_2$ is the fiber dispersion.\\
And the PSD of the NLI caused by jamming:
\begin{equation}
    G^{J}=\sum_{l\in r_i}N_l G_{l}^{J}
\end{equation}
with,

\begin{multline}
    G_{l}^{J}(f_m) = \phi G_m  \frac{\epsilon^2 + 2\epsilon P}{\Delta_m^2} \sum_{\substack{m' \in M_J \\ m' \neq m}}\text{ln}(\frac{f_{m,m'} + \Delta f_{m'}/2}{f_{m,m'} - \Delta f_{m'}/2}) 
\end{multline}
To ensure and maintain the QoT of connections, the control plane check if the value of SNR is higher than an SNR threshold according to the standard provided by the Service Level Agreement (SLA). Any connection with a SNR below the SNR threshold is not accepted. 

\section{Jamming Detection and Avoidance}\label{sec:detection}
\begin{figure}[!h]
    \begin{center}
        \includegraphics[width=0.4\textwidth]{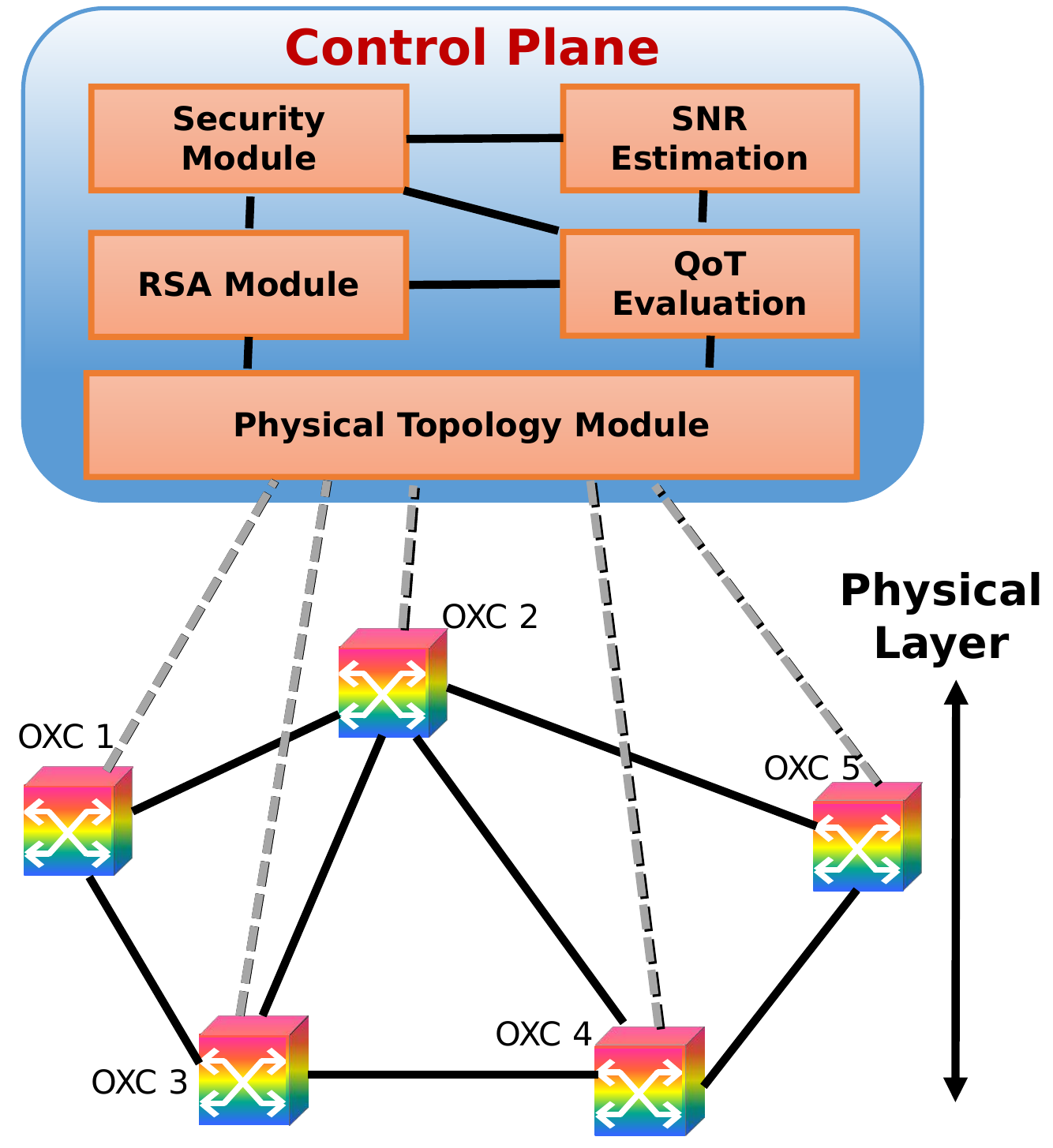}
        \caption{Overall architecture of the optical network control plane.}
        \label{fig:cp}
    \end{center}
\end{figure}

The overall architecture of the studied control plane is shown in Fig. \ref{fig:cp}. All the information about the state of the network are sent  to the \textit{Physical Topology Module}.

The \textit{RSA Module} finds the allocation solution based on the information provided by the\textit{Physical Topology Module}. After that, the solution is evaluated by the \textit{QoT Evaluation} module. The \textit{SNR estimation}, as its name says measures the SNR value based  on the route information. The obtained evaluations are sent to the \textit{Security Module}, which is responsible on ensuring the network security.

\begin{figure}[!h]
    \begin{center}
        \includegraphics[width=0.4\textwidth]{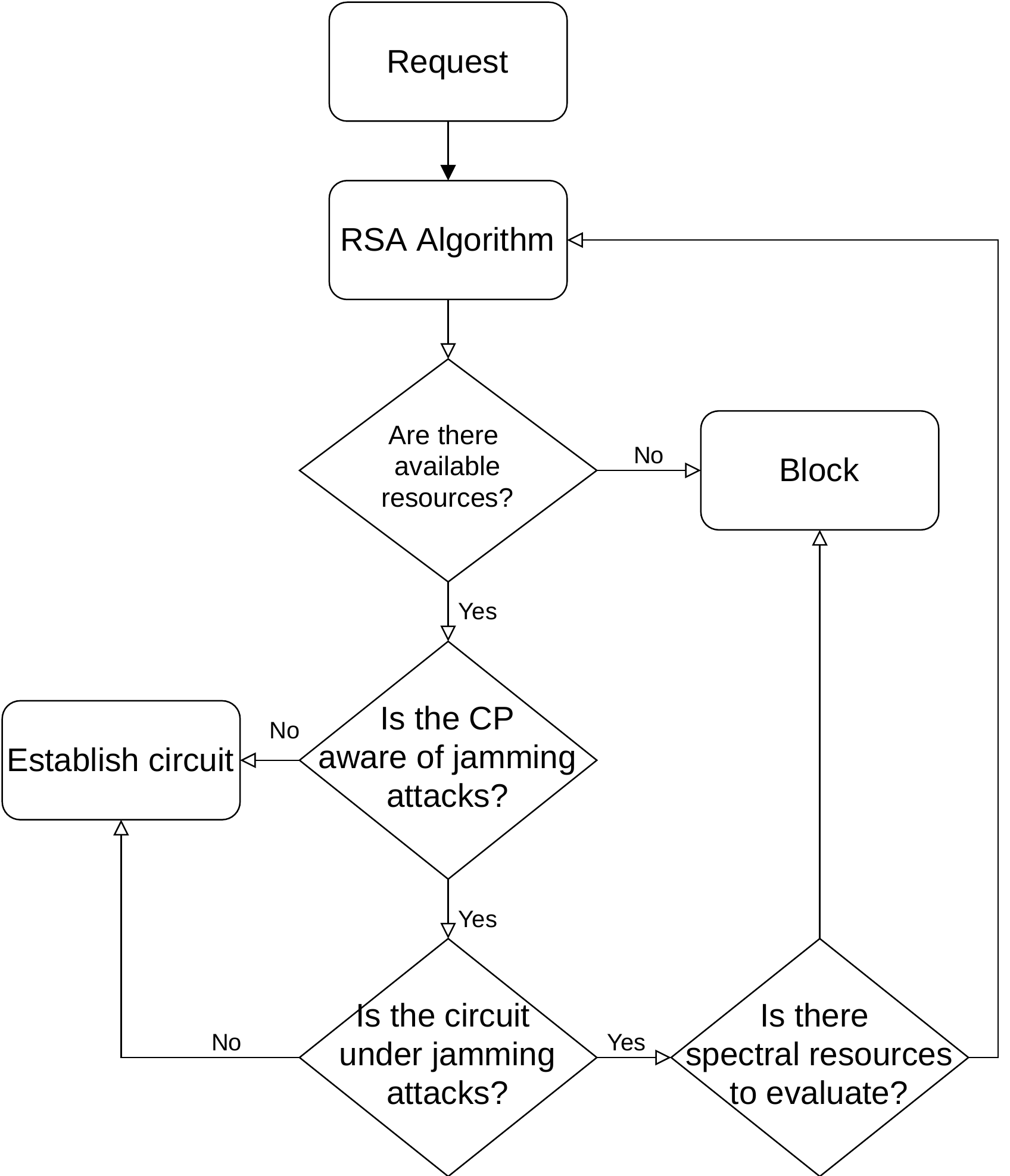}
        \caption{Flowchart for the control plane evaluation.}
        \label{fig:flowchart}
    \end{center}
\end{figure}

The control plane architecture is a key element to take advantage of many capabilities of an EON. In \cite{lopez:2018}, the authors consider the use of an interface, called TAPI \cite{tapi:2016}, which creates abstractions of some network functionalities. Among them is mentioned the ability to obtain information about physical impairments inside the network links to physical impairment computation models \cite{lopez:2018}. With this information, it is possible to propose a control plane aware of the jamming occurrence. In this study, a control plane model that prevents the creation of jammed circuits is evaluated. The goal is to verify the reduction of the jamming impact, mainly in the circuits that suffer from  out-of-band attacks. Figure \ref{fig:flowchart} shows the flowchart for answering or blocking calls operated by the control plane.

In the first step, the incoming circuit request is evaluated by the RSA algorithm, which is an extension of the control plane and can be any RSA algorithm found in the literature that suits the scenario. The RSA assesses the source and destination node pair to define resources available for the new circuit. The circuit request is blocked if no resources are found. Otherwise, the selected resources are informed to the control plane.

The control plane evaluates the selected solution and decides if the resources attend some quality restrictions. The control plane verifies: if the selected route respects the distance threshold allowed by the selected modulation level, if the nodes have enough transmitters and receivers for the new circuit, if the signal has acceptable noise power (in case of physical-impairment evaluation) and other measurements.

In this jamming-aware control plane proposal, the control plane uses the TAPI interface \cite{lopez:2018} to measure the power of new circuits and compare it with the estimated power for that circuit,  besides performing the quality evaluations mentioned previously. The SNR equation \cite{johannisson2014modeling} is used to estimate the power, and the circuit setting is the input. If any difference is found between the measured power and the estimated power, the control plane assumes that the circuit is crossing a jammed channel, and demands another set of spectral resources to the RSA algorithm. If no new resources are found, the request is blocked.

This proposed control plane model avoids the occurrence of in-band jamming, as no circuit is established inside the jammed channel. Furthermore, the out-of-band jamming intensity is reduced, and the only jamming interference is descendant from the empty jammed channel. The next section presents a performance evaluation from the jamming-aware control plane proposal, compared to the unaware control plane.

\section{Performance Evaluation} \label{sec:evaluation}
In this section, we analyze the performance of the proposed jamming-aware control plane on online network scenarios. In the simulation, the jamming is launched in three fixed slot ranges (50 to 59, 140 to 149 and 230 to 239), inside the 320 slot range. The reason of fixing the position of the attacker is to study the impact of in-band jamming on the chosen slots, and out-of-band jamming along the neighboring slots by looking to slots utilization. The effect of jamming can be propagated to other links through the circuits that cross the jammed slots in the attacked link.

\begin{figure}[!h]
    \begin{center}
        \includegraphics[width=0.4\textwidth, page=1]{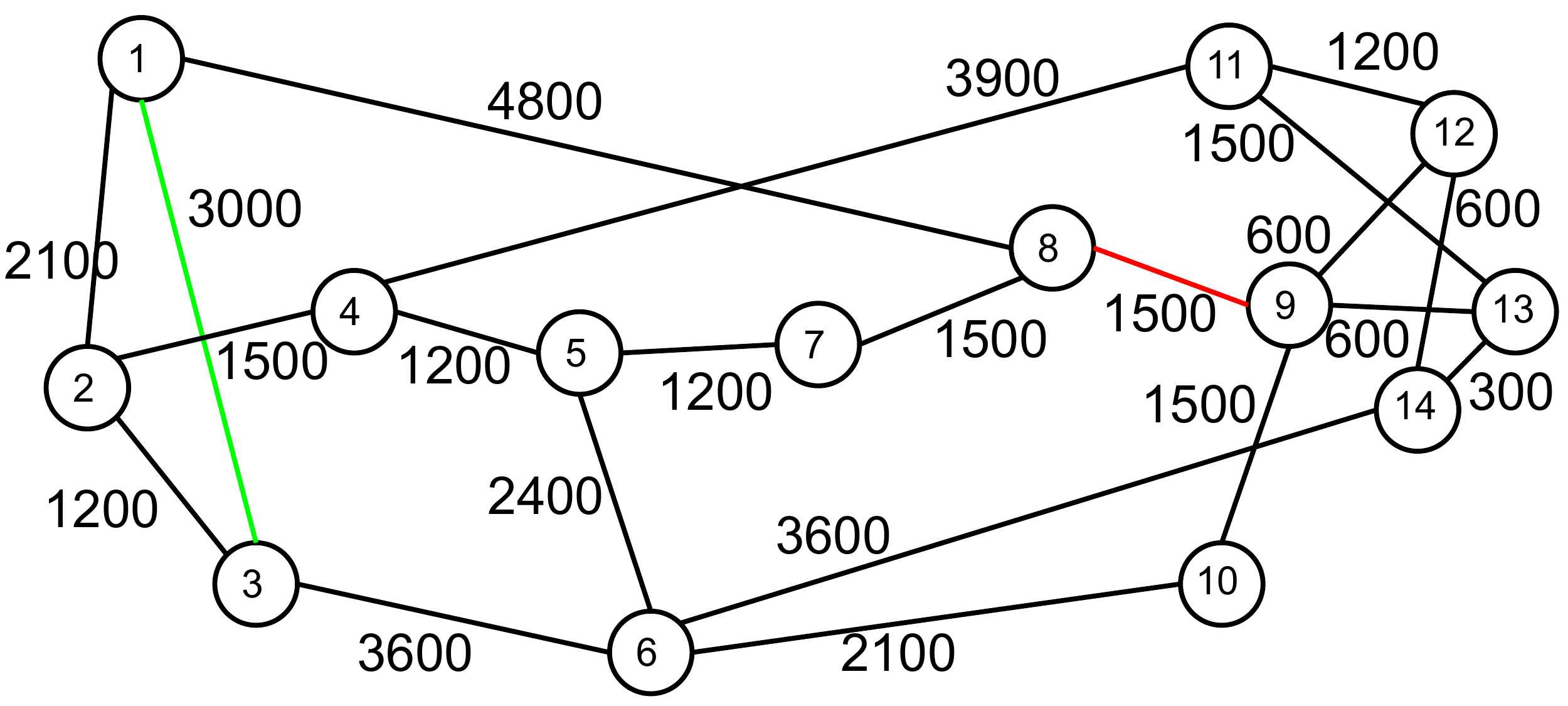}
        \caption{NSFNet topology, with the jammed link in red (links length in km).}
        \label{fig:topologies}
    \end{center}
\end{figure}

Two different scenarios are evaluated separately, and the jamming is applied in different links. In the first scenario, the attack occurs in the most used link. For the second scenario, the jamming is applied in the least used link. Simulations are performed previously, all with the same RSA algorithm, to measure the link utilization in terms of spectral reserves. Thus, the most used link is the one with highest average utilization in the whole simulation, and the least used is the link with lowest average utilization. The purpose is to evaluate the difference of impact in two different link profiles. Figure \ref{fig:topologies} shows the topology used in the simulations, the distance of each link in km, and the most (in red) and least (in green) used links.

Simulations are made with ONS simulator \cite{ONS}. In both scenarios, we consider links with 320 slots for each direction. All scenarios are evaluated under a fixed load of 200 Erlangs, and the results are the average of 10 replications, with 100,000 circuit requests each, following a Poisson distribution with mean holding time of 600 seconds, with negative exponential distribution and uniformly-distributed among all nodes-pairs. We use 3 different values of bandwidth: 40, 200 and 400 Gbps, uniformly selected. To solve the routing and spectrum assignment (RSA) problem, we adopt the basic Dijkstra algorithm for routing, and the \textit{First Fit}  policy for  slot assignment. The guardband between two adjacent lightpaths is assumed to be 2 slots. The circuits may use any of the modulation levels: BPSQ, QPSK, 8QAM, 16QAM, 32QAM and 64QAM, with $\text{SNR}_{th}$ equals to 9, 9, 12, 15, 18 and 21 respectively \cite{fontinele2017efficient}. The parameters for the SNR calculation are shown in table \ref{tab:physical} \cite{fontinele2017efficient}:

\begin{table}[!h]
\centering
\caption{Parameters for SNR calculation.}
\label{tab:physical}
\begin{tabular}{|c|c|}
\hline
\textbf{Variable}   & \textbf{Value}                \\ \hline
$P_{TX}$            & 0 dB                          \\ \hline
$\Delta f$          & 12.5 GHz                      \\ \hline
$\alpha$            & 0.2 dB/km                     \\ \hline
L                   & 100                           \\ \hline
$\gamma$            & $1.22 Wkm^{-1}$               \\ \hline
$\beta_2$           & $16 ps^2/km$                  \\ \hline
$v$                 & $1.93 \times 10^{14} Hz$      \\ \hline
$F$                 & $6\;dB$                        \\ \hline
\end{tabular}
\end{table}

\begin{figure*}[ht]
	\vspace{-2em}
	\centering
	\subfigure[Blocking probability without jamming avoidance.]
	{
		\includegraphics[width=0.47\textwidth]{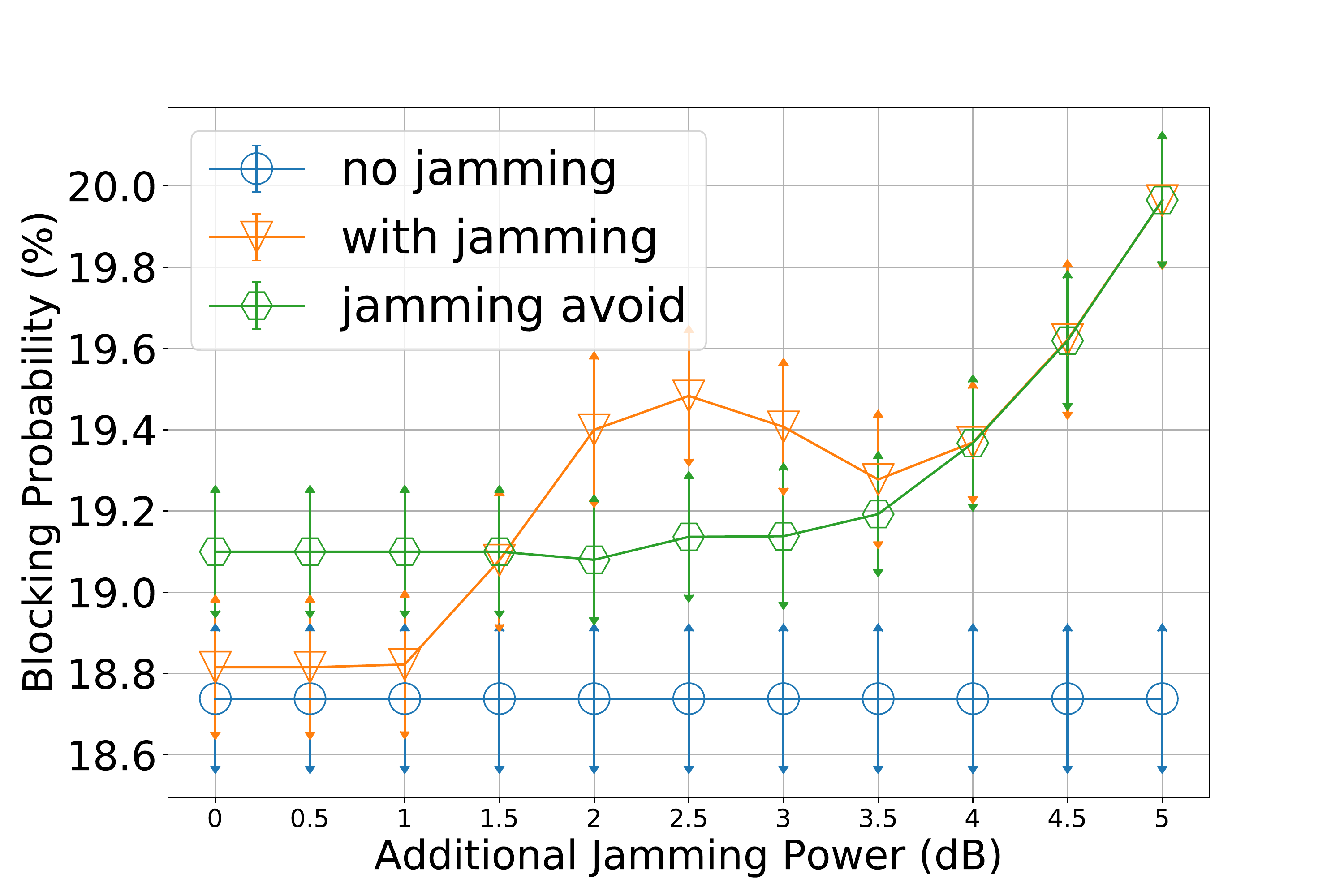}
		\label{fig:a}
	}
	\hfill
	\subfigure[Blocking probability with jamming avoidance.]
	{
		\includegraphics[width=0.47\textwidth]{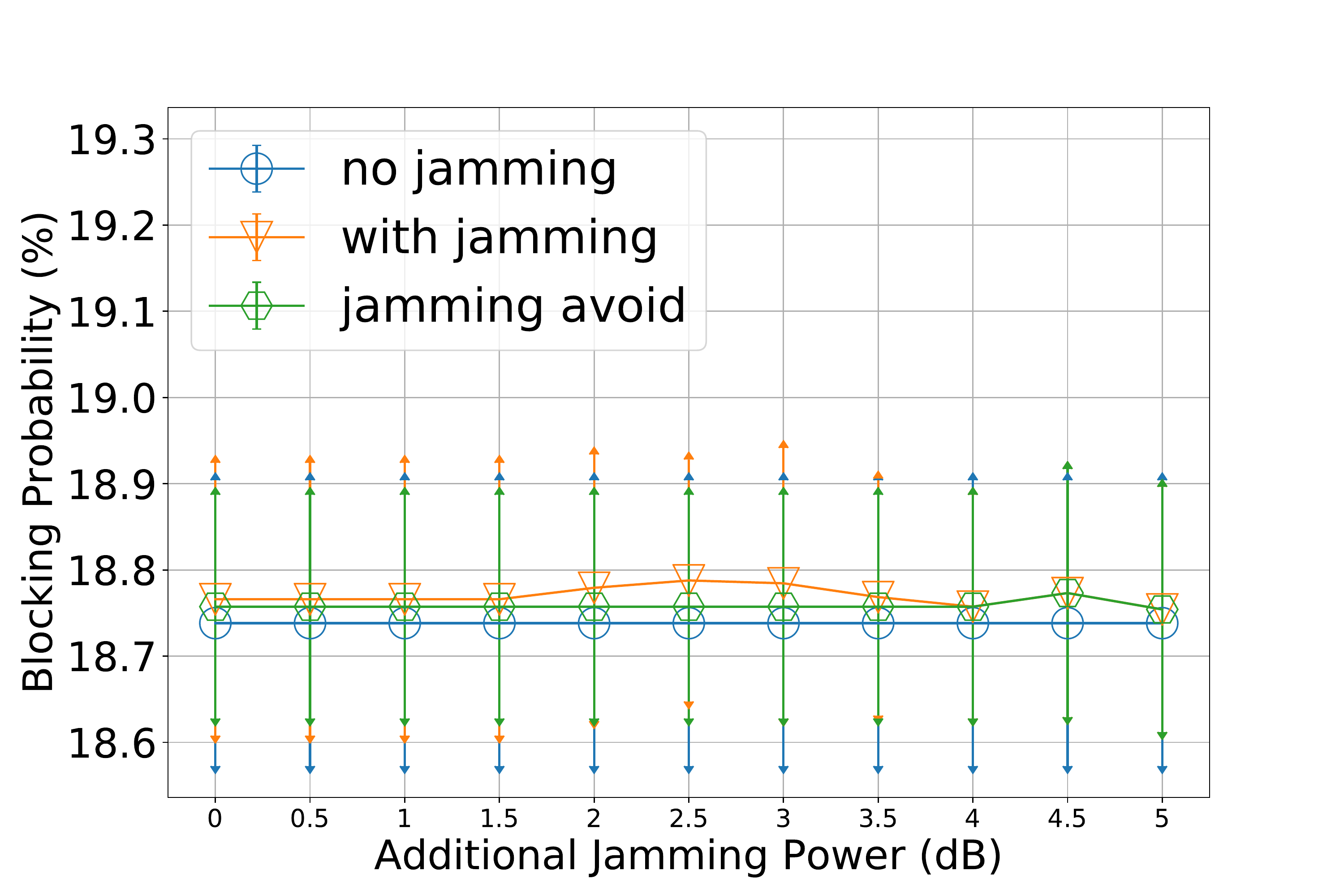}
		\label{fig:b}
	}
	\caption{Blocking probability for jamming attacks in the most used (a) and least used (b) links.}  
	\label{fig:blocking}
\end{figure*}

The jamming power $\epsilon$ varies from 0 to 5 dB, in a step of 0.5 dB. Higher values of $\epsilon$ (beyond 5 dB) are not much common in the EON jamming literature. Three different evaluations are performed. First, is assessed the blocking probability for \textit{i) scenarios without jamming}, \textit{ii) scenario with jamming attacks and unaware of it}, and \textit{iii) scenario with jamming attack and a jamming-aware control plane}. These evaluations are conducted for the jamming attack in the most used and the least used link. The second evaluation is also blocking probability, but performed in a smaller step of jamming variation (0.25 instead of 0.5), to show details in the range with high jamming impact. This comparison shows how different can be the impact in the network, depending on the attacked link. The third evaluation demonstrates the average slot utilization in the most used link, to clarify how different jamming approaches can affect the spectrum. Figure \ref{fig:blocking} presents the blocking probability for jamming attack in the most used (a) and least used (b) links.

To represent two different scenarios, the following notation is used: $MU$ is a reference to the \textit{most used link} and $LU$ is the \textit{least used link}. For the jamming occurrences, the $J$ is used to represent the scenario under jamming attack and unaware of it, and the $JA$ represents the jamming-aware control plane scenario. These representations are combined, like $MU-JA$ is used to represent the jamming-aware control plane in the scenario where the attack occurs in the most used link.

The blocking probability for the not-jammed scenario is constant in (a) and (b), and the $x$ axis represents the jamming power variation $\epsilon$ in the other scenarios. For the $MU-J$ scenario, the impact of small $\epsilon$ is indistinguishable, and the small increase of power almost do not impair the active circuits. However, as the jamming power increases, the blocking grows until reach the first peak, between scenarios of $\epsilon$ equals to 2 and 3 dB. In this peak is found the highest number of circuits under in-jamming attack. These are the circuits allocated inside the jammed channel. At this point, there are a lot of jammed circuits with SNR near the threshold, and these circuits prevent the allocation of new circuits.

For highest values ( $\epsilon$ greater than 3 dB), the creation of circuits inside the jammed channel is reduced. It occurs because the high power affects the SNR of jammed circuits spectrally close to each other, and the control plane do not allow the creation of circuits with SNR below the chosen modulation threshold, even in the unaware of jamming scenario. Thus, for higher values of $\epsilon$, the in-band jamming is reduced (as less circuits are created inside the jammed channel), and the out-of-band jamming grows, as the empty jammed channel still interferes in spectrally near circuits.

In the $MU-JA$ scenario, the blocking probability is slightly increased for low $\epsilon$ values. It occurs because some slots are avoided, even if the jamming interference is very low. The jammed channel is kept empty, and it keeps the blocking probability in a constant behavior until values of $\epsilon$ near $3.5$ dB. After this point, the out-of-band effect of the empty jammed channel is similar to $MU-J$ scenario, and the out-of-band effect in both scenarios is inevitable.

\begin{figure}
 \centering
   \includegraphics[scale=0.25]{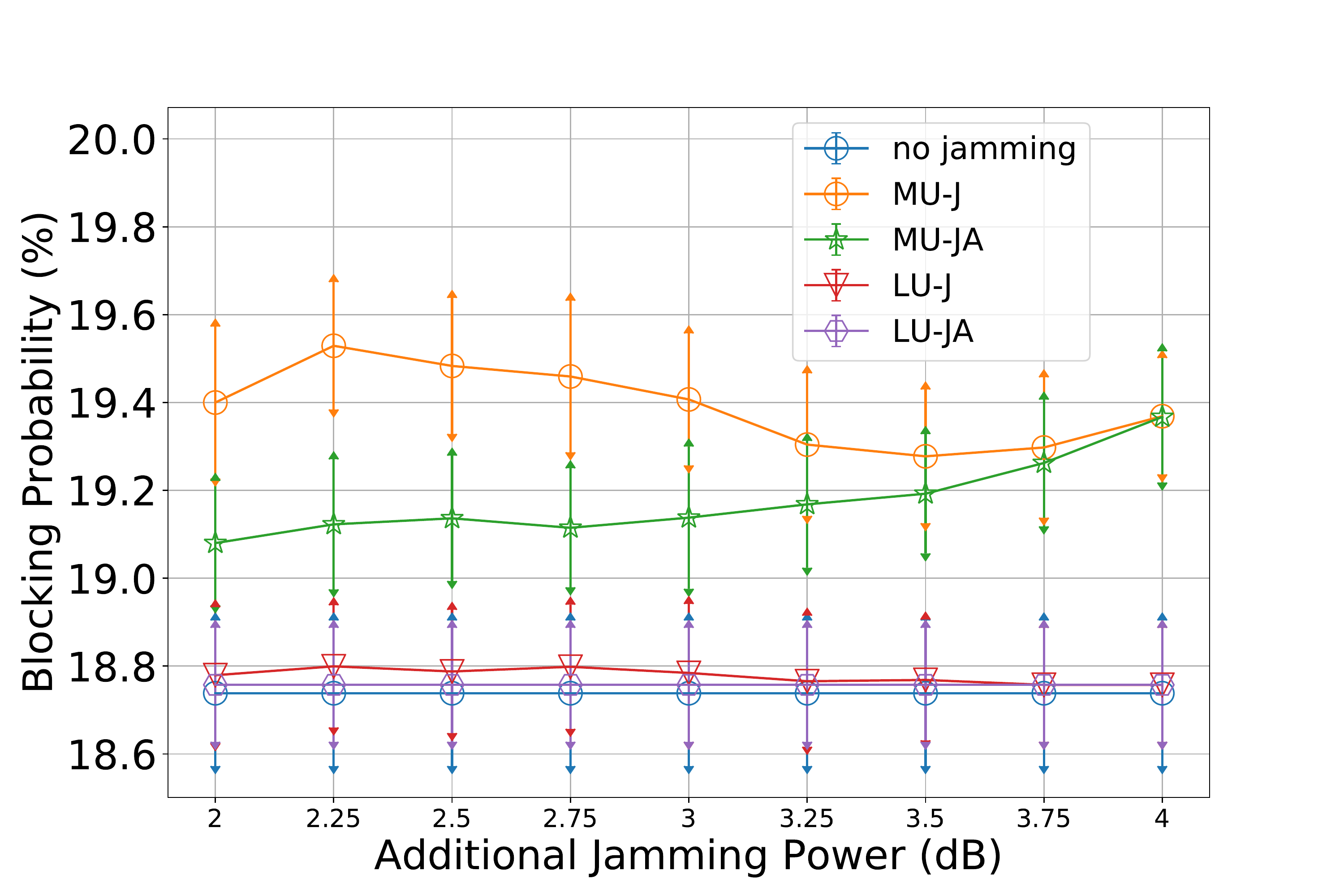}
 \caption{Blocking probability for jamming attacks in the most used (a) and least used (b) links, with zoom for the highest jamming impact.}
\label{fig:blockingZoom}
\end{figure}

For the simulations of Fig.\ref{fig:blocking}(b), the difference between the $LU-J$, $LU-JA$ and no jamming scenarios is almost negligible. This means that the jamming effect in a link with low usage has a small interference in the whole network performance. Figure \ref{fig:blockingZoom} shows the blocking probability with a smaller $\epsilon$ variation in the first blocking peak.

The results in Fig.~\ref{fig:blockingZoom} allows to perceive the different impacts of jamming attack in $MU$ and $LU$ links. In $LU$ scenario, even the worst case $LU-J$ has performance similar to the scenario without jamming. The $MU-J$ scenario achieves the worst blocking. In this $\epsilon$ range, with most intense in-band jamming, the jamming-aware control plane performs better, because it keeps the jammed channel empty and avoids the creation of in-band jammed circuits, which would spread the out-of-band jamming. It is found that the peak occurs to $\epsilon$ near $2.25$ dB.

\begin{figure*}
 \centering
   \includegraphics[scale=0.35]{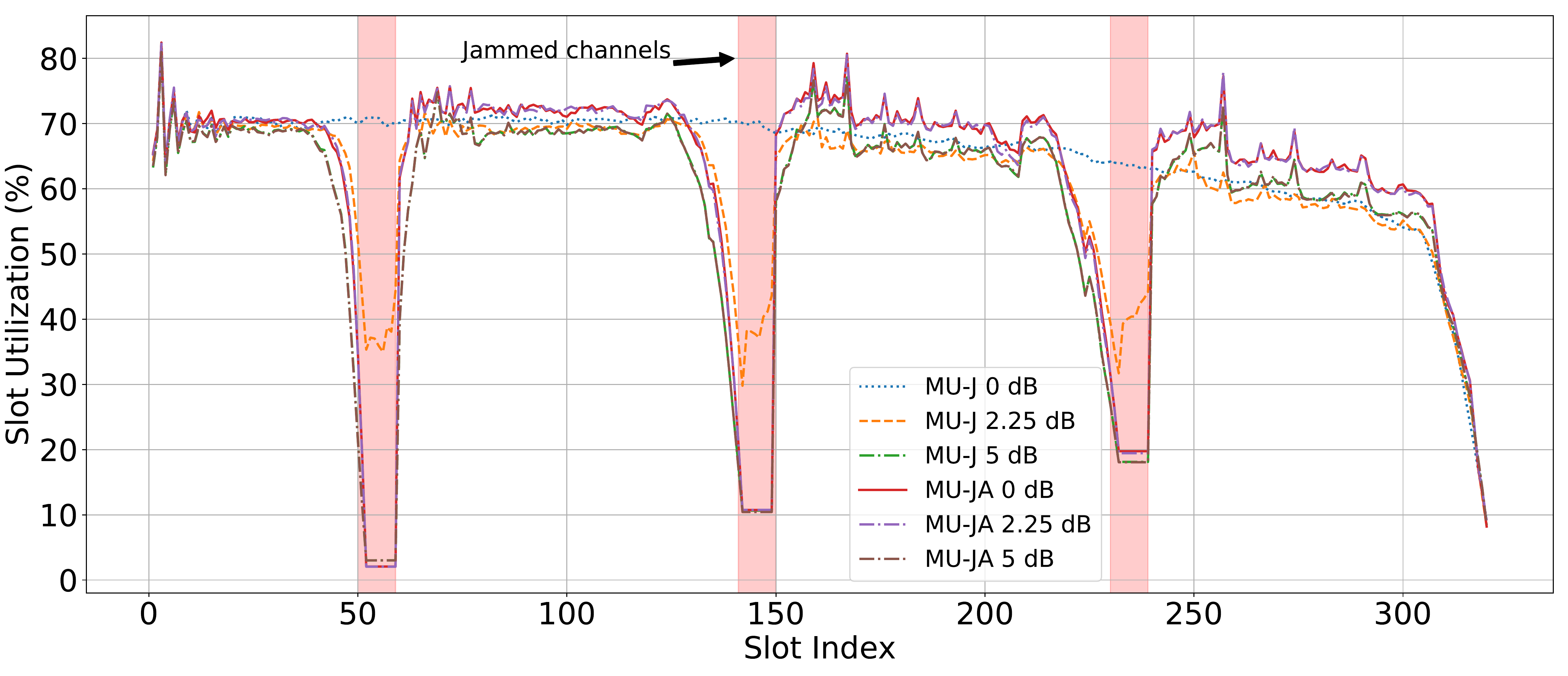}
 \caption{Average slot utilization of all links in the $MU$ scenario.}
\label{fig:slotUsage}
\end{figure*}

To demonstrate the behavior of circuits inside the optical spectrum, Figure \ref{fig:slotUsage} shows the average spectral utilization of all links in the network in $MU$ scenario.

For the $MU-J$ with 0 dB (equivalent to no jamming scenario), the slot utilization slowly decreases as the slot index increases. This behavior is a consequence of First Fit allocation policy, in which slots are allocated from the first index to the last. For the other scenarios, some gaps are included, and this can occur in two cases: as a consequence of an increasing in-band jamming, which decreases the slot utilization in the jammed channel when $\epsilon$ increases ($MU-J$ 2.25dB and $MU-J$ 5 dB cases); or as a consequence of jamming avoidance by the control plane, when the jammed slots are always free and avoided by the control plane (in $MU-JA$ 0.0dB, $MU-JA$ 2.25 dB and $MU-JA$ 5 dB cases).

The presented results shows that the link selected by the attacker plays an important role in the attack intensity. Furthermore, the jamming power $\epsilon$ may be adjusted to a value not so high and still can cause great impact in the network. These results open some research opportunities, like the study of different jamming strategies, variations in size of jammed channel and the application of the jamming model in other types of optical networks.
 
\section{Conclusions} \label{sec:conclusion}

In this paper, we analyzed jamming power attacks in elastic optical networks, by embedding the insertion of jamming power into the physical layer model. We proposed an architecture of a jamming aware control plane. Then, we studied the impact of in-band and out-of-band jamming on the blocking probability and slots utilization into two different scenarios, considering the jamming occurrence in the most used and the least used link. 

As a future work, we plan to incorporate techniques of jamming detection in the proposed security module, to provide a precise information about the jammer position, as well as proposing accurate prevention mechanism.

\bibliographystyle{./IEEEtran}
\bibliography{./mybib}

\end{document}